# Prevention of Metro Rail Accidents and Incidents in Stations Using RFID Technology

Farshid Sahba
Institute for Informatics and Automation Problems
National Academy of Sciences of Armenia
Yerevan, Armenia
f_sahba@yahoo.com

Ramin Sahba
Department of Electrical and Computer Engineering
The University of Texas at San Antonio
San Antonio, TX, USA
ramin.sahba@utsa.edu

*Abstract*— Today, metro is one of the urban infrastructure and plays an important role in urban transport. The safety and health of people in a city are always important, and transport in the metro should also be safe. When subway trains operate, it is possible to occur various accidents such as an exit from the rails or collision with a possible obstacle on the rails (human or another train). In this paper, a model is proposed based on RFID technology in which the train is equipped with a RFID reader, and a control circuit with a microcontroller as well as placing RFID active tags at specific points of the path. When the train approaches the tagged points of the route, the train tag reader scans the tag number, and then the microcontroller identifies the status of the environment by retrieving the information from the internal database. Then, the control circuit adjusts the rotational speed of the electric motor and the train speed consequently based on that data. In this way, the speed of the train is adapted automatically and promptly to the conditions and the probability of an accident in unforeseen circumstances is reduced.

## I. Introduction

Today, metro trains are among the most important and most used vehicles in urban transport. Every day, millions of people in metropolitan cities around the world use metro, and the safety of this large population is very important when using metro. From time to time, there are horrific incidents in metro stations or metro lines, and a system that can reduce these incidents and increase the safety of metro passengers is very useful and essential.

In the near future, the railway industry will rely more on transportation systems equipped with smart technologies [1]. These smart technologies will bring more safety to the trains and the passengers. Radio Frequency Identification (RFID) and Wireless Sensor Networks (WSN) are parts of these technologies. RFID helps to detect presence and location of objects, and WSN helps to sense and monitor the environment [2]. [3] describes a WSN architecture for secure railways. The proposed system identifies spoilage on railways using acceleration measurement and ultrasounds. Researchers in [4] propose a system for early earthquake detection based on WSN for use in high-speed railways. [5] presents another system that utilizes electromagnets and image processing to detect objects on the railroads. [6-8] use RFID to secure and control some other systems.

In this paper, we present a model in which several types of RFID readers and RFID tags have been used to reduce the frequency of station and rail accidents. According to official news and reports, there have been several terrible railroad incidents in the world. Table 1 shows the number of rail accidents, by type of accident, in Europe in 2016 [9]. The model presented in this article has been developed for dealing with common types of train incidents including collision with human, collision with another train, and running off the rails (derailment) [10-14].

Table 1. Number of rail accidents, by type of accident, in Europe in 2016 [9]

| | TOTAL | Collisions | Derailments | Level crossing accidents (incl. pedestrians) | Accidents to persons by rolling stock in motion (excl. suicides) | Fires in rolling stock | Other accidents |
|---|---|---|---|---|---|---|---|
| EU-28 | 1 787 | 101 | 68 | 433 | 1 069 | 36 | 80 |
| Belgium | 22 | 2 | 0 | 12 | 7 | 0 | 1 |
| Bulgaria | 40 | 3 | 6 | 5 | 24 | 2 | 0 |
| Czech Republic | 87 | 6 | 2 | 34 | 38 | 2 | 5 |
| Denmark | 5 | 0 | 0 | 1 | 3 | 1 | 0 |
| Germany | 310 | 29 | 3 | 50 | 183 | 6 | 39 |
| Estonia | 15 | 3 | 0 | 8 | 1 | 1 | 2 |
| Ireland | 0 | 0 | 0 | 0 | 0 | 0 | 0 |
| Greece | 13 | 1 | 0 | 1 | 10 | 0 | 1 |
| Spain | 45 | 6 | 7 | 10 | 22 | 0 | 0 |
| France | 146 | 7 | 5 | 48 | 79 | 4 | 3 |
| Croatia | 23 | 0 | 2 | 5 | 15 | 0 | 1 |
| Italy | 99 | 4 | 2 | 15 | 72 | 1 | 5 |
| Latvia | 18 | 0 | 0 | 3 | 15 | 0 | 0 |
| Lithuania | 20 | 0 | 1 | 6 | 13 | 0 | 0 |
| Luxembourg | 2 | 1 | 0 | 1 | 0 | 0 | 0 |
| Hungary | 162 | 1 | 2 | 27 | 115 | 9 | 8 |
| Netherlands | 28 | 7 | 0 | 7 | 11 | 1 | 2 |
| Austria | 87 | 7 | 5 | 31 | 41 | 0 | 3 |
| Poland | 265 | 5 | 16 | 76 | 168 | 0 | 0 |
| Portugal | 38 | 4 | 5 | 8 | 21 | 0 | 0 |
| Romania | 184 | 1 | 0 | 42 | 140 | 1 | 0 |
| Slovenia | 11 | 1 | 1 | 8 | 0 | 0 | 1 |
| Slovakia | 60 | 2 | 1 | 12 | 42 | 2 | 1 |
| Finland | 18 | 4 | 1 | 6 | 4 | 0 | 3 |
| Sweden | 36 | 2 | 3 | 7 | 15 | 4 | 5 |
| United Kingdom | 53 | 5 | 6 | 10 | 30 | 2 | 0 |
| Channel Tunnel | 0 | 0 | 0 | 0 | 0 | 0 | 0 |
| Norway | 16 | 11 | 2 | 0 | 3 | 0 | 0 |
| Switzerland | 35 | 2 | 0 | 2 | 26 | 1 | 4 |
| Former Yugoslav Republic of Macedonia | 88 | 1 | 16 | 14 | 12 | 0 | 45 |
| Turkey | 120 | 6 | 23 | 51 | 36 | 0 | 4 |

## II. Modeling

The subway train is a vehicle on the rails, and the most important parameter for controlling this vehicle is its speed. This speed is between zero and 200 km/h, and the speed parameter must be changed appropriately at the station, the curves, the straight path, the slopes, and while facing possible obstacles [15-18]. The speed is directly proportional to the motor rotational speed, so by controlling the rotational speed of the electric motor and, therefore, the vehicle speed, it is possible to control the train to prevent possible accidents [19,20].

Fortunately, locomotives of metro trains have electric motors, so it is possible to control the rotational speed of the electric motor using a control circuit equipped with a processor [21]. The processor must decide on the received data from the environment and set the rotational speed of the electric motor based on that. In the proposed model of this paper, we install RFID active tags in the physical environment of the metro system, especially the hazardous points of it, and equip the front part of the train with an RFID reader to receive data from the environment. The RFID reader then, when the train moves on the rails and approaches the hazardous points of the path, reads the number in the tag and sends it to the processor. The processor finds out the status of the environment by accessing data in a predefined database stored in the ROM, and adjusts the rotational speed of the electric motor based on that. Hazardous points include curves, stations, slopes, and obstacles (humans or other trains). RFID active tags will be installed on the tunnel walls, curves, slopes, the stations and the exterior part of the body of the rear wagons. For the passengers, the tag can be placed in the mobile phone, smart watch, smart card or wristband. In Figure 1, the status of the rails, the train, and the tags in a curve, the station, and confrontation with a possible obstacle are schematically shown.

In addition to essential hardware infrastructure, this system requires well designed software to manage and control the situation [22-29]. The software is implemented and run in the system processor and determine how the processor commands control the electric motor of the train.

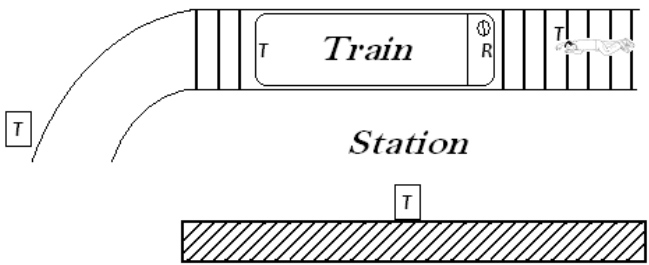

Fig. 1. status of the rails, the train, and the tags in a curve, the station, and confrontation with a possible obstacle

In the first algorithm, when the train reaches sloping routes (uphill or downhill), the rotational speed of the electric motorincreases or decreases based on the conditions. The pseudocode of this algorithm is shown in Figure 2.

```
void  Slope_Resolver()
{
    cin>> Tag_Code;
    Slope_Condition = Restore(Tag_Code,
     DataBase);
    cin>> Motor_Revolution;
    while (!(Is_Suitable(Motor_Revolution,
        Slope_Condition)))
    {
        if (Slope_Condition > 0)
            ++Motor_Revolution;
        else
            --Motor_Revolution;
        cin>> Tag_Code;
        Slope_Condition = Restore(Tag_Code,
         DataBase);
    }
}
```

Fig. 2. Pseudocode of the algorithm for controlling train electric motors based on slope of the route.

In the second algorithm of this system, the train is controlled in a way that in right turns or left turns of the path, rotation speed of the right wheels or left wheels can be reduced depending on the conditions. Figure 3 shows the pseudocode of this algorithm.

```
void  Bend_Resolver()
{
    cin>> Tag_Code;
    Bend_Condition = Restore(Tag_Code,
     DataBase);
    cin>> Right_Wheel_Revolution >>
     Left_Wheel_Revolution;
    while (!(
     (Is_Suitable(Right_Wheel_Revolution,
     Bend_Condition)) &&
     (Is_Suitable(Left_Wheel_Revolution,
     Bend_Condition))))
    {
        if (Bend_Condition > 0)
            --Right_Wheel_Revolution;
            // Turn to Right
        else
            --Left_Wheel_Revolution;
            // Turn to Left
        cin>> Tag_Code;
        Bend_Condition = Restore(Tag_Code,
         DataBase);
    }
}
```

Fig. 3. Pseudocode of the algorithm for controlling rotation speed of the right and left wheels of the train based on the turns in the path.

In the third algorithm of the system, the train will slow down its speed and stop when reaching the station, or confronting obstacles such as another train or a human on the rails. Like previously defined algorithms, this algorithm also starts with reading the tag code. The pseudocode of this algorithm is shown in Figure 4.

```
void Stop_Train()
{
    cin>> Tag_Code;
    Stop_Condition = Restore(Tag_Code,
    DataBase);
    cin>> Train_Speed;
    while (!(Is_Suitable(Train_Speed,
        Stop_Condition)))
    {
        --Train_Speed;
        cin>> Tag_Code;
        Stop_Condition = Restore(Tag_Code,
        DataBase);
    }
}
```

Fig. 4. Pseudocode of the algorithm for slowing down the train speed and stopping it based on observed conditions.

Formulas for determining the suitability of train speed, rotational speed of the right wheel or the left wheel, and the rotational speed of the electric motor on the slope that are used in the *Is_Suitable()* procedure are based on the train weight, distance to the barrier or station, route slope, and the route curve as well as the power of the motor and its current rotational speed [30,31]. Calculation of these formulas needs technical knowledge in the field of mechanics which is out of scope of this paper with an IT view to the problem.

It should be mentioned that the distance to the hotspots or obstacles can be calculated from the intensity of the received radio signals [32]. Also, the detection of the rotational speed of the electric motor and the weight of the train requires appropriate sensors.

### III. EXPERIMENT AND RESULTS

To evaluate the effectiveness of a model, specific test scenarios are needed, and different models require different methods [33-43]. To test the effectiveness of the designed model, the program was coded in Code Vision at the mechatronics lab of Rayan Shid Nama Eng. Co. (RSN) and was loaded onto the Atmega16 microcontroller. Then, the microcontroller was installed on the interface circuit board with command relays. The interfacing circuit was placed on a small model train with two electric motors (one on each side of the train), and the wiring was done in such a way that the relays could command electric motors. Then the train was placed on a laboratory railway track, which was somewhat similar to the actual metro rails, and moved on. Figure 5 shows the small model train on its railway track in the station.

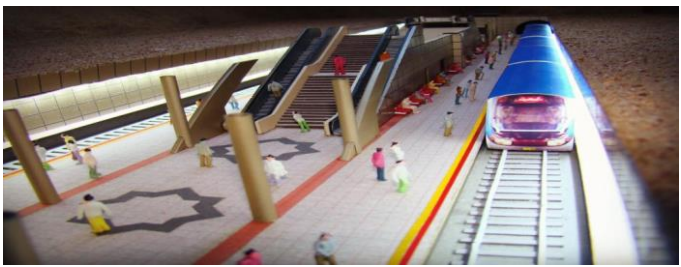

Fig. 5. The model train in the station

In first experiment, the implemented system based on the proposed method was activated on the model train. 60 trips were carried out on replica rails, 16 of which had an accident-prone situation (human obstacle, other trains, or sharp turn), which resulted in a total of 58 safe trips and only 2 incidents. The same experiment was retested while the implemented system was deactivated. Again, out of 60 trips, 16 of which had a dangerous situation. The outcome included 46 safe trips and 14 incidents.

However, it should be mentioned that in an actual situation, due to the presence of a train driver, the probability of a disaster will be relatively less in both cases. It is because the human agent's alertness in the control of the train has a very impactful effect on its safety and preventing incidents.

### IV. CONCLUSION

With the experiments conducted in the implemented model, we can say that the efficiency of the proposed system in this article was relatively effective to the reduction of disasters. However, for the definitive conclusion, this system should be installed on a pilot basis on urban subways and further investigated.

The strengths of this system are the simplicity, affordability and availability of needed hardware and software to launch it. However, the implementation of this system needs installing of required equipment at the metro stations and tracks, which takes time and is fairly expensive.

The system proposed in this paper can be developed to be used for safety of transportation in urban streets, road transportation, carrying materials and workers in mines, transporting of materials, products and equipment in agricultural fields, as well as the safety of the transportation in inter-city trains.